\documentclass[twocolumn]{aastex631} 

\received{}
\revised{}
\accepted{}

\shorttitle{assembly history for central galaxies}
\shortauthors{Lyu et al. 2023}

\graphicspath{{./}{figures/}} 

\begin{document}

\title{From Halos to Galaxies. VII. The connections between stellar mass growth history, quenching history and halo assembly history for central galaxies}
\correspondingauthor{Yingjie Peng}
\email{yjpeng@pku.edu.cn}
	
\author{Cheqiu Lyu}
\affiliation{Department of Astronomy, School of Physics, Peking University, 5 Yiheyuan Road, Beijing 100871, People’s Republic of China}
\affiliation{Kavli Institute for Astronomy and Astrophysics, Peking University, 5 Yiheyuan Road, Beijing 100871, People’s Republic of China}

\author{Yingjie Peng}
\affiliation{Department of Astronomy, School of Physics, Peking University, 5 Yiheyuan Road, Beijing 100871, People’s Republic of China}
\affiliation{Kavli Institute for Astronomy and Astrophysics, Peking University, 5 Yiheyuan Road, Beijing 100871, People’s Republic of China}

\author[0000-0002-4534-3125]{Yipeng Jing}
\affiliation{Department of Astronomy, School of Physics and Astronomy, Shanghai Jiao Tong University, Shanghai 200240, People’s Republic of China}
\affiliation{Tsung-Dao Lee Institute, and Shanghai Key Laboratory for Particle Physics and Cosmology, Shanghai Jiao Tong University, Shanghai 200240, People’s Republic of China}

\author[0000-0003-3997-4606]{Xiaohu Yang}
\affiliation{Department of Astronomy, School of Physics and Astronomy, Shanghai Jiao Tong University, Shanghai 200240, People’s Republic of China}
\affiliation{Tsung-Dao Lee Institute, and Shanghai Key Laboratory for Particle Physics and Cosmology, Shanghai Jiao Tong University, Shanghai 200240, People’s Republic of China}

\author[0000-0001-6947-5846]{Luis C. Ho}
\affiliation{Kavli Institute for Astronomy and Astrophysics, Peking University, 5 Yiheyuan Road, Beijing 100871, People’s Republic of China}
\affiliation{Department of Astronomy, School of Physics, Peking University, 5 Yiheyuan Road, Beijing 100871, People’s Republic of China}

\author[0000-0002-7093-7355]{Alvio Renzini}
\affiliation{INAF--Osservatorio Astronomico di Padova, Vicolo dell’Osservatorio 5, I-35122 Padova, Italy}

\author[0000-0002-6137-6007]{Bitao Wang}
\affiliation{Kavli Institute for Astronomy and Astrophysics, Peking University, 5 Yiheyuan Road, Beijing 100871, People’s Republic of China}

\author[0000-0002-3775-0484]{Kai Wang}
\affiliation{Kavli Institute for Astronomy and Astrophysics, Peking University, 5 Yiheyuan Road, Beijing 100871, People’s Republic of China}

\author{Bingxiao Xu}
\affiliation{Kavli Institute for Astronomy and Astrophysics, Peking University, 5 Yiheyuan Road, Beijing 100871, People’s Republic of China}

\author{Dingyi Zhao}
\affiliation{Department of Astronomy, School of Physics, Peking University, 5 Yiheyuan Road, Beijing 100871, People’s Republic of China}
\affiliation{Kavli Institute for Astronomy and Astrophysics, Peking University, 5 Yiheyuan Road, Beijing 100871, People’s Republic of China}

\author[0000-0002-6961-6378]{Jing Dou}
\affiliation{School of Astronomy and Space Science, Nanjing University, Nanjing 210093, People’s Republic of China}

\author[0000-0002-3890-3729]{Qiusheng Gu}
\affiliation{School of Astronomy and Space Science, Nanjing University, Nanjing 210093, People’s Republic of China}

\author[0000-0002-4985-3819]{Roberto Maiolino}
\affiliation{Cavendish Laboratory, University of Cambridge, 19 J.J. Thomson Avenue, Cambridge, CB3 0HE, UK}
\affiliation{Kavli Institute for Cosmology, University of Cambridge, Madingley Road, Cambridge, CB3 0HA, UK}
\affiliation{Department of Physics and Astronomy, University College London, Gower Street, London WC1E 6BT, UK}

\author[0000-0002-4803-2381]{Filippo Mannucci}
\affiliation{INAF--Osservatorio Astrofisico di Arcetri, Largo Enrico Fermi 5, I-50125 Firenze, Italy}

\author[0000-0003-3564-6437]{Feng Yuan}
\affiliation{Key Laboratory for Research in Galaxies and Cosmology, Shanghai Astronomical Observatory, Chinese Academy of Sciences, 80 Nandan Road, Shanghai 200030, People’s Republic of China}
\affiliation{University of Chinese Academy of Sciences, No. 19A Yuquan Road, Beijing 100049, People’s Republic of China}

\begin{abstract}
		The assembly of galaxies over cosmic time is tightly connected to the assembly of their host dark matter halos. We investigate the stellar mass growth history and the chemical enrichment history of central galaxies in SDSS-MaNGA. We find that the derived stellar metallicity of passive central galaxies is always higher than that of the star-forming ones. This stellar metallicity enhancement becomes progressively larger towards low-mass galaxies (at a given epoch) and earlier epochs (at a given stellar mass), which suggests strangulation as the primary mechanism for star formation quenching in central galaxies not only in the local universe, but also very likely at higher redshifts up to $z\sim3$. We show that at the same present-day stellar mass, passive central galaxies assembled half of their final stellar mass $\sim 2$ Gyr earlier than star-forming central galaxies, which agrees well with semi-analytic model. Exploring semi-analytic model, we find that this is because passive central galaxies reside in, on average, more massive halos with a higher halo mass increase rate across cosmic time. As a consequence, passive central galaxies are assembled faster and also quenched earlier than their star-forming counterparts. While at the same present-day halo mass, different halo assembly history also produces very different final stellar mass of the central galaxy within, and halos assembled earlier host more massive centrals with a higher quenched fraction, in particular around the ``golden halo mass'' at $10^{12}\mathrm{M_\odot}$. Our results call attention back to the dark matter halo as a key driver of galaxy evolution. 
		
\end{abstract}


	~\\
	
	\section{Introduction}

	In the past decades, a concordance model termed $\Lambda$CDM has been established which is remarkably successful in reproducing the large-scale structures. In this framework, galaxies harbor and co-evolve with the dark matter halos, which make up the majority of the gravitational potential. It is believed that the properties of galaxies such as stellar mass, star formation rate (SFR), metallicity, and etc are likely to be influenced by the mass growth rate of the host dark matter halo, or the rate of merger events \citep{Wechsler2018, Bose2019}. Since dark matter is invisible, if such interplay between baryon and dark matter can be statistically established, we are able to gain insight into properties of dark matter halo through baryonic probes. However, the detail of the galaxy-halo connection still remains elusive.

	On the halo side, cosmological N-body simulations have shown that the clustering of halos is not solely determined by their mass but also other secondary properties, such as their formation time \citep{Wechsler2006, Gao2007, Zentner2014, Lim2016, Contreras2019, Mansfield2020}. Galaxies in halos with same present-day mass but different assembly history might have drastically different properties, which is termed as halo assembly bias. Though the halo assembly bias was first discovered in high-resolution numerical simulations in 2005, it has proven to be challenge to identify it in observations, largely due to the difficulty in accurately measuring the secondary parameter. Currently, there is no convincing observational evidence for the halo assembly bias. Some studies have claimed that the bias signal is significant \citep{Yang2006, Cooper2010, Miyatake2016}, while others have argued that the signal is weak or the statistical fluke \citep{Blanton2007, Dvornik2017, Tinker2017, Tinker2018, Salcedo2020}.
	
	On the baryon side, similarly, one of the core aim in galaxy evolution is to understand how galaxies assemble their stellar mass. The star formation history (SFH) is a crucial property to understand the assembly of the stellar mass. Traditionally, the information of SFH was extracted by the photometry based SED fitting and stellar population synthesis modeling. With the ever-increasing usage of the spectra data, more sophisticated codes have been developed to perform full spectral fitting of observed spectra, with varying focuses on emission lines, stellar kinematics, stellar population, or all-in-one approaches. Examples include pPXF \citep{Cappellari2004, Cappellari2017}, STARLIGHT \citep{CidFernandes2005}, VESPA \citep{Tojeiro2007}, ULySS \citep{Koleva2009}, FIREFLY \citep{Wilkinson2017}, and PROSPECTOR \citep{Johnson2021}. Combination of data from integrated field spectroscopy (IFS) with single stellar population (SSP) decomposition enables the acquisition of not only global but also spatially resolved information about how galaxies assemble their stellar mass and metal content. For instance, the fossil record method has been successfully applied to modern IFS galaxy surveys like CALIFA \citep{Perez2013, Garcia-Benito2017, LopezFernandez2018} and MaNGA \citep{Ibarra-Medel2016, Goddard2017, Sanchez2018, Sanchez2019} to reconstruct the local and global SFH or stellar mass growth history, and the chemical enrichment history of galaxies \citep[see, e.g.][]{Ibarra-Medel2016, Fraser-McKelvie2019, Peterken2019, Peterken2020}.
	
	Another related puzzle is how the process of assembly ceases, or how the galaxies shut oﬀ their star formation activity (quenching). Current consensus is that the processes of quenching can be broadly divided into two categories \citep[e.g.][]{Kauffmann2003a,Balogh2004,Peng2010}: mass quenching, which is internally driven processes that operate in both central and satellite galaxies; environment quenching, which is externally driven processes that operate mostly in satellite galaxies. Various of physical mechanisms have been proposed: for mass quenching, the candidates include AGN feedback \citep{Croton2006,Fabian2012}, morphological quenching \citep{Martig2009}, gravitational quenching \citep{Genzel2014}, angular momentum quenching \citep{Peng2020, Renzini2020} and etc.; and for environment quenching, including strangulation \citep{Larson1980,Balogh2000,Peng2015}, ram pressure stripping \citep{Gunn1972, Abadi1999, Quilis2000}, tidal stripping and harassment \citep{Farouki1981,Moore1996}, major merger \citep{Mihos1996, Hopkins2008}, halo quenching \citep{Dekel2006} and etc. Though equipped with various options, a definitive understanding of quenching is still lacking.
	
	The observation of dark matter is expensive and challenging. Currently, there are only a few reliable samples of measurements to the mass of the dark matter halo using strong (weak) lensing and HI kinematic. On the other hand, baryonic implementation in N-body simulations is expensive and time-consuming as well. In this sense, semi-analytic models (SAMs) is an economic way to study the linkage between the galaxies and halos. It is less time-consuming on fine-tuning the parameters than N-body simulations; and it also implements with the recipes of formation and evolution of dark matter halo from N-body simulations, and simple analytic descriptions of all relevant baryonic physics, including the heating and cooling of gas, the star formation, and the merging of galaxies, etc. Therefore, SAMs provides us a useful and straightforward tool to explore the halo assembly history and galaxy growth history, simultaneously. Comparison between the observations and results from SAMs would gain us insight about connection between galaxy and halo and their evolution. In this work, we compare the stellar mass growth history of central galaxies in SDSS-MaNGA with L-GALAXIES and investigate the connection between stellar mass growth history, quenching history and host halo assembly history for central galaxies. We describe the adopted data and sample selection in Section \ref{data}, followed by an exploration of how and whether the galaxy growth is affected by dark matter halos in Section \ref{results}. In Section \ref{discussion}, we discuss our results in the context of previous works and highlight the limitations and prospects of our study. Finally, we summarize our work and present our conclusions in Section \ref{conclusion}. Throughout this paper, we adopt a \citet{Chabrier2003} initial mass function (IMF) and use the following cosmological parameters: $\Omega_{\mathrm{m}}= 0.3$, $\Omega_{\mathrm{\Lambda}} = 0.7$, and $H_0=70$ km s$^{-1}$ Mpc$^{-1}$.
	
	\section{Data}\label{data}
	
	\subsection{MaNGA}
	
	MaNGA (Mapping Nearby Galaxies at Apache Point Observatory) is one of the three core programs of the fourth generation of SDSS \citep[SDSS-IV,][]{Blanton2017}, which includes $\ge10,000$ galaxies at $0.01 \lesssim z \lesssim 0.15$ with spatially resolved, high-quality spectra ($\lambda/\Delta \lambda\sim 2,000$) \citep[see][]{Yan2016, Wake2017}. The main sample is constructed with a flat distribution in stellar mass, and down to $\sim 10^9M_{\odot}$. The size of integral field unit (IFU) varies from 12 arcsec to 32 arcsec, which is sufficiently large to cover most of the selected galaxies. The spectra coverage spans from 3600 to 10300 Angstrom, which enables reliable constraints on the parameters such as stellar mass, star formation history, and stellar metallicity, etc. 
	
	In this work, we adopt the MaNGA {\ttfamily Pipe3D\_v3\_1\_1} catalog \citep{Sanchez2022}, which was produced via the Pipe3D pipeline that was designed to utilize IFU to investigate the properties of stars and ionized gas \citep{Sanchez2016}. The parameters in the catalog were computed from a full spectrum fitting by exploring series of stellar population synthesis models in grid of ages and metallicities. Specifically, Pipe3D analyzes the data cubes by sampling them linearly in wavelength and preprocessing them to account for Galactic extinction and a normalized spectral resolution. In their analysis, contiguous spaxels are grouped together to create a spatial tessella bin to achieve high signal-to-noise ratio (S/N). Pipe3D applies a simple stellar population fitting method to model the stellar continuum. The SSP spectrum is adjusted based on the mean stellar velocity and multiplied by the Cardelli extinction curve \citet{Cardelli1989} to correct for MW-like dust extinction. Additionally, it is convolved with a Gaussian function to account for stellar velocity dispersion. For the SSP fitting, Pipe3D utilizes the {\ttfamily MaStar\_sLOG} stellar library from \citet{Yan2019}. This library consists of 273 SSP spectra covering 39 ages ranging from 1 million years to 13.5 billion years, and 7 metallicities spanning from 0.006 $Z_\odot$ to 2.353 $Z_\odot$. The initial mass function of \citet{Salpeter1955} is adopted to fit each MaNGA cube. Consequently, the derived properties are obtained by fitting the observed galaxy spectrum to a combination of stellar population synthesis models with different ages and metallicities. In order to construct the stellar mass growth history and chemical enrichment history, Pipe3D employs a non-parametric approach by decomposing the history into individual star formation bursts generated by SSPs. Additionally, it considers a group of stars that originated from the same gas at a specific time with a particular metal content \citep[e.g.][]{Panter2007, ValeAsari2009, Perez2013, Ibarra-Medel2016}. The outputs of Pipe3D include the integrated properties for 10,226 galaxies, such as main stellar population properties, emission lines, and kinematics, etc, to serve the subsequent scientific purposes. Although this method can accurately reproduce the observed details of spectra, it is prone to significant uncertainties \citep[e.g.][]{CidFernandes2014, Sanchez2016, Sanchez2020}. The primary source of uncertainty stems from the fossil record method employed by Pipe3D, which uses combinations of stellar ages and metallicities from different lookback times for each galaxy to fit the luminosity evolution obtained from spatially resolved IFS observations. Consequently, there may be inconsistencies in the stellar mass growth and chemical enrichment history of individual galaxies.
	
	In this study, we select galaxies with stellar mass $M_* \ge 10^9M_{\odot}$ to ensure galaxies with well-measured properties, as recommended by Pipe3D. We select galaxies with spectral fitting quality control flag \texttt{QCFLAG}($=0$: good; $=2$: bad; $>2$: warning) equals 0 as advised to exclude galaxies with wrong redshift or low S/N. In addition, MaNGA utilized the stellar mass function from nearby SDSS galaxies \citep{Blanton2017} to estimate the volume accessible in each bin of stellar mass, redshift and color. We then perform the volume correction for galaxies with reliable $V_\mathrm{max\_w}$ values. Moreover, the integrated star formation rate (SFR) was derived from dust corrected H$\alpha$ luminosity \citep{Kennicutt1998}. For the stellar mass growth history and chemical enrichment history, we use ``{\ttfamily Tx}'' (lookback time at which the galaxy has formed x\% of its stellar mass) and ``{\ttfamily ZH\_x}'' (stellar metallicity, normalized to the solar value, in logarithmic-scale at the time at which x\% of the current stellar mass was formed) as provided by the catalog \citep{Sanchez2016, Sanchez2018}. Finally, we limit the galaxies in our sample as central galaxies to avoid the complexity introduced by the environmental effects. The selection was performed by cross-matching our sample with the group catalog in \citet{Yang2007} where the central galaxies are the most massive in each group. A final sample of 5,706 galaxies makes the selection cuts.
		
	\subsection{L-GALAXIES}\label{LG}
	
	L-GALAXIES is a semi-analytic model that uses dark matter halo merger trees to trace the formation and evolution of galaxies based on sets of empirical relations \citep{Guo2011, Henriques2015}. We utilize the version implemented in \citet[][hereafter H15]{Henriques2015} in this work, which is implemented on the subhalo merger trees that were constructed by the Millennium simulations \citep{Springel2005}, scaled to the first-year Planck cosmology \citep{PlanckCollaboration2014}. The simulation box has a side length of $500h^{-1}\mathrm{Mpc}$ and each dark matter particle has a mass of $8.6\times10^{8}h^{-1}\mathrm{M_\odot}$. Halos are identified using the friends-of-friends (FoF) method \citep{Davis1985}, whereas subhalos are captured by the SUBFIND algorithm \citep{Springel2001}. Since the subhalo merger trees are constructed by identifying the unique descendant for each subhalo \citep{Springel2005}, we are able to trace the main branch by recursively identifying the most massive progenitor subhalo. In addition, the most massive subhalo in each FoF halo is designated as the main halo, with the galaxy attributed to it is the central galaxy, while the remaining subhalos and galaxies are designated as satellites.
	
	One of the main advantage of this implemented version is that it properly accounts for the baryon cycle in the process of galaxy evolution. In this model, the baryon is initially accreted onto dark matter halos in terms of hot gas, and then cools down to cold gas (atomic and molecular hydrogen) via radiation, and forms stars. During the lifetime of the stars, a fraction of baryon was return to the ISM in the form of stellar wind or supernova feedback. Furthermore, galaxy mergers can trigger starbursts and facilitate the formation and growth of central massive black holes. Central black holes are assumed to grow primarily through the accretion of cold gas during mergers, but also through quiescent accretion from the hot atmosphere. The latter process generates jets and bubbles, which in turn produce radio mode feedback. This feedback suppresses cooling onto the galaxy, eliminates the supply of cold gas, and subsequently quenches star formation. When a dark matter halo falls into a larger halo, it becomes a satellite subhalo, and the galaxies within it become satellite galaxies that suffer from environmental effects such as tidal and ram-pressure stripping. All these processes have been properly treated based on sets of empirical recipes and parameter fine-tune to calibrate with the observations. L-GALAXIES provides well-matched distributions of stellar mass, color, specific star formation rate, and luminosity-weighted stellar ages across the stellar mass range $8.0 \leq \log M_{*} / \mathrm{M_\odot} \leq 12.0$ \citep{Henriques2015}. Furthermore, L-GALAXIES offers improved statistical significance due to its larger box volume.
	
	It is important to note that different simulations and semi-analytic models employ varying recipes, particularly in terms of feedback mechanisms, which may impact the results we have obtained. In this study, our results are based on the physical recipe implemented in the H15 version of L-GALAXIES. In this work, we use galaxies with stellar mass larger than $10^{9.0} \mathrm{M_\odot}$. It should be noted that due to the resolution limit of the dark matter particles in the Millennium simulations, galaxies with stellar masses more massive than $10^{9.5} \mathrm{M_\odot}$ are more reliable to use, as recommended by \citet{Henriques2015}.
	
	\section{Results}\label{results}
	\subsection{Stellar mass growth history in MaNGA}
	
	\begin{figure}[htb]
		\centering
		\includegraphics[width=0.45\textwidth]{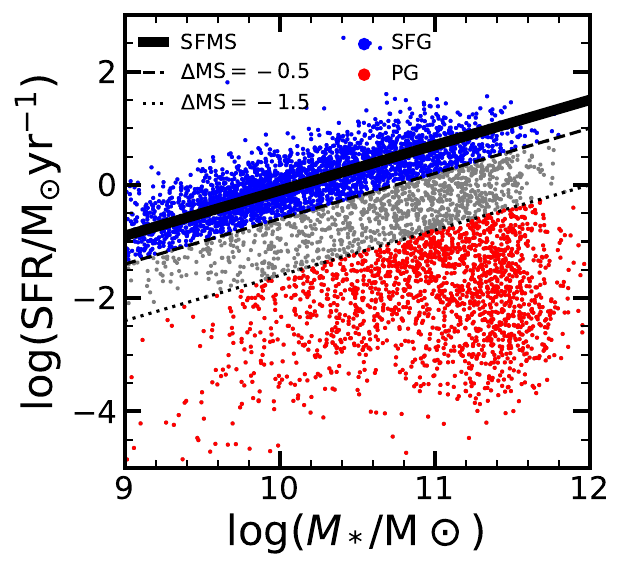}
		\caption{Stellar mass and star formation rate for MaNGA  sample. The thick black solid line indicates the Star Forming Main Sequence (SFMS), and the thin dashed and dotted lines represent the downward offset of 0.5 and 1.5 dex from the SFMS, respectively. The galaxies with $\Delta\mathrm{MS}>-0.5$ are classified as SFGs (blue dots), whereas those with $\Delta\mathrm{MS}<-1.5$ are classified as PGs (red dots). The galaxies that fall between the two thin lines are green valley galaxies (gray dots), which are not the focus of this study.
		\label{M-SFR}}
	\end{figure}

	\begin{figure*}[htb]
		\centering
		\includegraphics[width=1\textwidth]{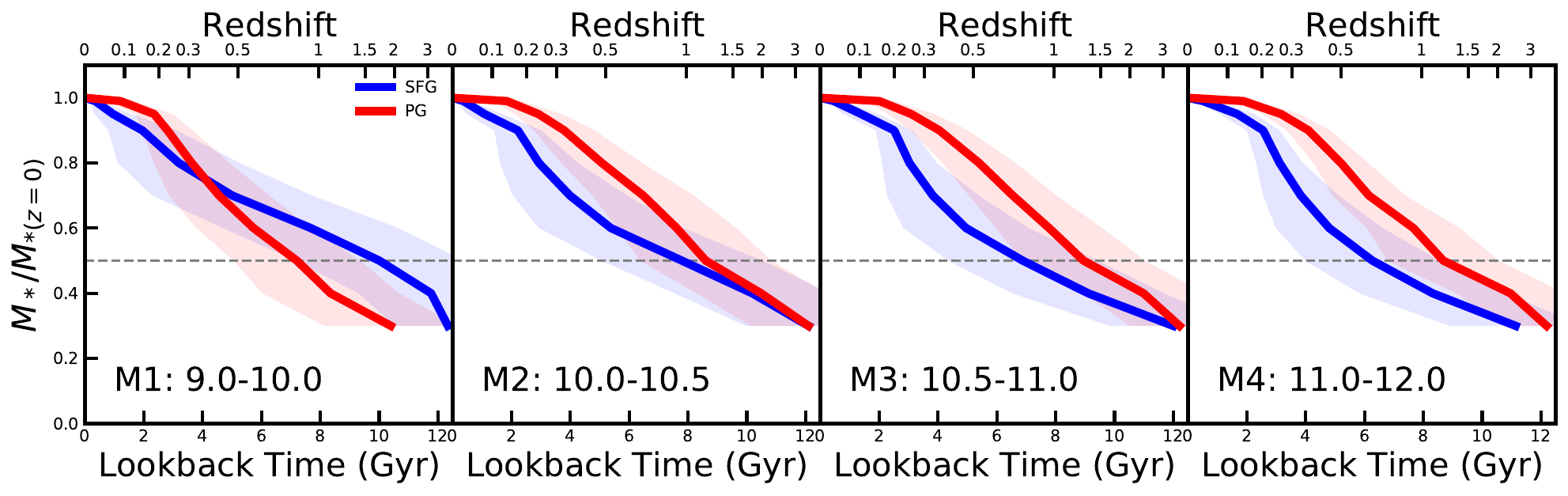}
		\caption{Stellar mass growth history of star-forming central galaxies (SFGs, blue) and passive central galaxies (PGs, red) for MaNGA sample, normalized to the stellar mass at $z=0$. Each panel corresponds to a different range of present-day stellar mass: $\log(M_*/\mathrm{M_\odot})=9.0-10.0$ (M1), $10.0-10.5$ (M2), $10.5-11.0$ (M3), and $11.0-12.0$ (M4), respectively. The solid curves depict the median growth history, with shaded regions representing $1\sigma$ error on the median value in the x-axis. Dashed lines mark half of the present-day stellar mass. \label{MassGrowth_MaNGA_rel}}
	\end{figure*}
		
	To investigate the stellar mass growth history and quenching history of central galaxies, we divide our sample into present-day star-forming galaxies (SFGs) and passive galaxies (PGs) based on $\Delta$MS, which is the offset distance in y-direction to the SFMS (Star Forming Main Sequence). The ridge line of SFMS is presented as a linear fitting as follows:
	\begin{equation}
		\log(\mathrm{SFR}/ \mathrm{M_\odot yr^{-1}})=0.8\log(M_*/\mathrm{M_\odot})-8.1,
	\end{equation} 
	\begin{equation}
		\Delta\mathrm{MS}=\log(\mathrm{SFR})-\log(\mathrm{SFR_{SFMS}}),
	\end{equation} 
	where galaxies with $\Delta\mathrm{MS}>-0.5$ are classified as SFGs, whereas those with $\Delta\mathrm{MS}<-1.5$ are classified as PGs. The galaxies that fall between them are green valley galaxies, which are not the focus of this study. With these boundaries, a total of 2,864 central SFGs and 1,769 central PGs remain. The distribution of our MaNGA sample on the $M_*$-SFR plane is shown in Figure \ref{M-SFR}. We have also tested alternative definition of the main sequence using a curved SFMS to define SFGs and PGs, and find small changes to the results presented in this paper. Unless otherwise specified, SFGs/PGs in this paper always refer to the star-forming/passive central galaxies.
	
	Figure \ref{MassGrowth_MaNGA_rel} illustrates the stellar mass growth history based on the fossil record method from MaNGA Pipe3D, normalized to present-day stellar mass. Panels from the left to right represent curves with four distinct ranges of present-day median stellar mass, respectively. At the same present-day stellar mass, the SFGs (blue) and PGs (red) have distinct stellar mass growth history. On average, the SFGs display an monotonically increase in their stellar mass through cosmic time, while the PGs exhibit more of a two-phases growth. At high redshift, PGs show a faster stellar mass growth compared to SFGs, and their pace of growth appear to slow down indicated by the flattening at the lookback time $\sim 2-4$ Gyr. As indicated by the gray dashed lines in Figure \ref{MassGrowth_MaNGA_rel}, most of PGs roughly assembled half of their final mass budget at $z \sim 1$, which is $\sim 2$ Gyr earlier than their star-forming counterparts. The only exception is for low-mass galaxies in the leftmost panel. It could be due to that observation uncertainties for low-mass galaxies are larger; or physically, the star formation history of low-mass galaxies is more bursty and susceptible to the process of rejuvenation, thus rendering the stellar mass growth history between SFGs and PGs indistinguishable. In the figures presented in this paper, we uniformly plot statistical error of each galaxy sample. It is worth noting that although we cannot neglect the uncertainties of evolution history of individual galaxy from full spectral fitting, these uncertainties have little impact on statistical analyses of galaxy populations. 
	
	It is important to note that the stellar mass growth history from MaNGA Pipe3D exhibits significant error along the x-axis, particularly at high redshift, which means that stellar mass growth has a great deal of uncertainties. With this in mind, it is hard to robustly tell galaxies in which stellar mass bin assembled the half of the present-day stellar mass earlier from solely based on the results of full spectral fitting. For instance, when comparing the stellar mass growth history derived from MaNGA Pipe3D with the theoretically extracted history in \citet{Yang2012}, we find that for massive galaxies, MaNGA Pipe3D reaches half of the present-day stellar mass later, whereas for low-mass galaxies, they reaches it earlier. Furthermore, in comparison to the observational results in \citet{Madau2014}, the stellar mass growth history derived from MaNGA Pipe3D appears to be shallower before $z\sim0.5$. Given the substantial uncertainties associated with the MaNGA Pipe3D-derived history, we must  acknowledge that our paper primarily focuses on the relative values rather than the absolute values, with particular emphasis on highlighting the trend differences between SFGs and PGs.
	
	\subsection{Chemical enrichment history in MaNGA}
	
	\begin{figure*}[htb]
		\centering
		\includegraphics[width=1\textwidth]{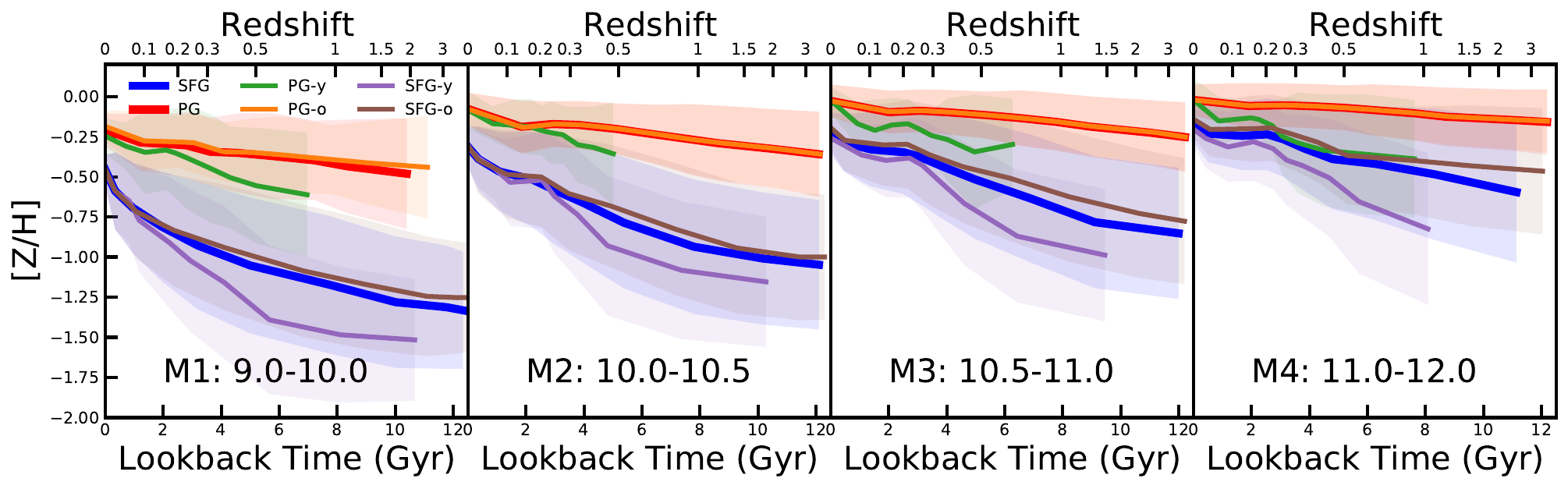}
		\caption{Chemical (stellar metallicity) enrichment history of star-forming central galaxies (SFGs, blue) and passive central galaxies (PGs, red) for MaNGA sample. Each panel shows the median values of galaxies in a different range of present-day stellar mass: $\log(M_*/\mathrm{M_\odot})=9.0-10.0$ (M1), $10.0-10.5$ (M2), $10.5-11.0$ (M3), and $11.0-12.0$ (M4), respectively. In each panel, the SFGs and PGs are further divided into young (age $<$ 4 Gyr: ``SFG-y", green / ``PG-y", purple) and old (age $>$ 4 Gyr: ``SFG-o", brown / ``PG-o", orange) subsamples based on their mass-weighted age in MaNGA Pipe3D catalog. The shaded regions represent $1\sigma$ error on the median value. \label{Zenrich_MaNGA_abs_agebin}}
	\end{figure*}

	\begin{figure}[htb]
		\centering
		\includegraphics[width=0.45\textwidth]{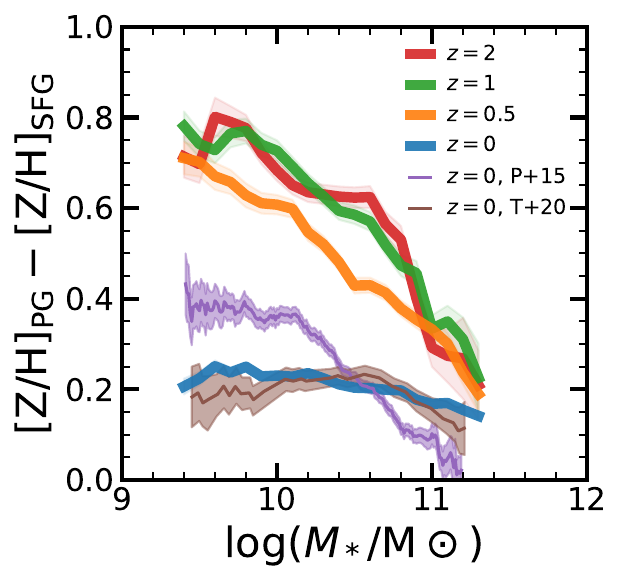}
		\caption{Stellar metallicity difference between passive central galaxies (PGs) and star-forming central galaxies (SFGs) as a function of stellar mass at $z \sim$ 0 (blue), 0.5 (orange), 1 (green), 2 (red) for MaNGA sample. The median values are calculated in the sliding box of 0.5 dex in stellar mass, and the shaded regions are errors with 1$\sigma$ uncertainty. The purple line and brown line indicate the stellar metallicity difference between passive galaxies and star-forming galaxies obtained in \citet{Peng2015} (labeled as ``P+15'') and \citet{Trussler2020} (labeled as ``T+20''), with shaded regions corresponding to the $1\sigma$ error on the mean value and median value, respectively. \label{MZRdifference_MaNGA}}
	\end{figure}

	Figure \ref{Zenrich_MaNGA_abs_agebin} presents the stellar metallicity enrichment history for SFGs (blue) and PGs (red) in four present-day stellar mass bins. Overall, the stellar metallicity of SFGs shows a gradual increase over time, with a slightly more rapid increase for low-mass galaxies. This is also consistent with the observed trends in stellar metallicity and gas-phase metallicity over time \citep[e.g.][]{Gallazzi2005, Maiolino2008, Mannucci2010}. The similarity of the trends between stellar metallicity and gas-phase metallicity is expected since the metal locked in stars originates from gas \citep{Peng2014}. Meanwhile, the measured stellar metallicity of PGs appears already saturated at high redshifts.
	
	Across the panels, the derived stellar metallicity of PGs is always higher than that of the SFGs at a given epoch and present-day stellar mass, and this stellar metallicity difference is larger for present-day low-mass galaxies. While at a given stellar mass, the stellar metallicity difference becomes increasingly larger at an earlier epoch. Bear in mind that the dichotomy of SFGs and PGs is based on the star formation statuses at present-day, and the PGs based on such classification could be SFGs at high redshift. To test whether this progenitor bias affects the increasing trends of difference in stellar metallicity with redshift, we further categorize the SFGs and PGs into two subsamples, respectively, based on their age (young with age $<$ 4 Gyr, and old with age $>$ 4 Gyr) as shown in thin colored lines in Figure \ref{Zenrich_MaNGA_abs_agebin}. For all present-day mass bins, the trend of stellar metallicity difference between SFGs and PGs remains similar for both young and old subsamples. Therefore, we conclude that the trends observed in main sample is not affected by the progenitor bias.
	
	With the mass growth history and chemical enrichment history, as shown in Figure \ref{MassGrowth_MaNGA_rel} and Figure \ref{Zenrich_MaNGA_abs_agebin}, we can derive the stellar metallicity difference between PGs and SFGs as a function of stellar mass at $z \sim 0, 0.5, 1$ and $2$, as shown in Figure \ref{MZRdifference_MaNGA}. We perform this because the relative value of stellar metallicity is much more reliable than the absolute value considering the uncertainties in Pipe3D. On the other hand, the stellar metallicity difference between PGs and SFGs can be used to discriminate efficiently between the gas removal and strangulation processes \citep{Peng2015}. As shown in Figure \ref{MZRdifference_MaNGA}, on average, the stellar metallicity difference is higher for low-mass galaxies than for massive galaxies at a given epoch. At a given stellar mass, the stellar metallicity difference increases with the redshift. We also overplot the stellar metallicity difference between SFGs and PGs in \citet{Peng2015} (labeled as ``P+15'' in purple) and \citet{Trussler2020} (labeled as ``T+20'' in brown), with shaded regions corresponding to the $1\sigma$ error on the mean value and median value, respectively. The stellar metallicity used in \citet{Peng2015} was from \citet{Gallazzi2005}, who studied the the spectral absorption features of galaxies from SDSS DR4; while stellar metallicity in \citet{Trussler2020} were derived from full spectral fitting using the code FIREFLY \citep[][]{Comparat2017, Goddard2017, Wilkinson2017}. The trend of difference in metallicity in this work is in broad agreement with that in \citet{Peng2015} and \citet{Trussler2020}.
	
	The trends in Figure \ref{MZRdifference_MaNGA} are consistent with the scenario of strangulation, which suggests PGs have higher metallicity than their star-forming counterparts at a given stellar mass \citep[see][]{Peng2015, Trussler2020}. \citet{Peng2015} showed that the enhancement in the stellar metallicity during the process of strangulation is mainly determined by their gas fraction (the timescale of strangulation is another important factor) in the sense that the higher the gas fraction is, the larger the enhancement. On the other hand, it is well-known that in general, the gas fraction of low-mass SFGs is higher than that of massive galaxies; and at a given stellar mass, the gas fraction of SFGs at higher redshift is higher than that at lower redshift \citep{Tacconi2018, Tacconi2020}. Therefore, the gas-rich SFGs with higher enhancement in metallicity difference at high redshift, are in line with the prediction of strangulation scenario. This implies that strangulation is not only the primary mechanism responsible for halting star formation activity in local central galaxies \citep{Peng2015}, but also likely to be an universal mechanism that works at higher redshifts.
	
	In addition to strangulation, the removal of gas (e.g. via feedback) is also required in quenching the most massive central galaxies. Figure \ref{MZRdifference_MaNGA} reveals that at $\log(M_*/\mathrm{M_\odot}) >11.0$, the stellar metallicity difference at high redshifts is not significantly different from that at $z \sim 0$. However, the molecular gas fraction at $z\sim 2$ is notably higher than local value \citep[e.g. more than twice as high at $\log(M_*/\mathrm{M_\odot}) \sim11.0$, according to][]{Tacconi2020}. Thus, some more violent processes that can quickly remove the gas may be required in the process of quenching for massive central galaxies at $z\sim2$ (e.g. AGN feedback).
	
	We caution that the trends depicted in Figure \ref{MZRdifference_MaNGA} is only plausible in a qualitative sense, since an significant increase in the stellar metallicity difference from $z=0.5$ to $z=0$ may be quantitatively unreasonable. This may be attributed to the inaccurate inference of the chemical enrichment history solely based upon the spectrum at $z \sim 0$. Direct measurements on the metallicity of galaxies at high redshift using spectroscopic instruments such as MOONS \citep[Multi-Object Optical and Near-infrared Spectrograph,][]{Cirasuolo2014} will be decisive.
	
	\subsection{Stellar mass growth and halo assembly history in L-GALAXIES}
	\begin{figure*}[htb]
		\centering
		\includegraphics[width=1\textwidth]{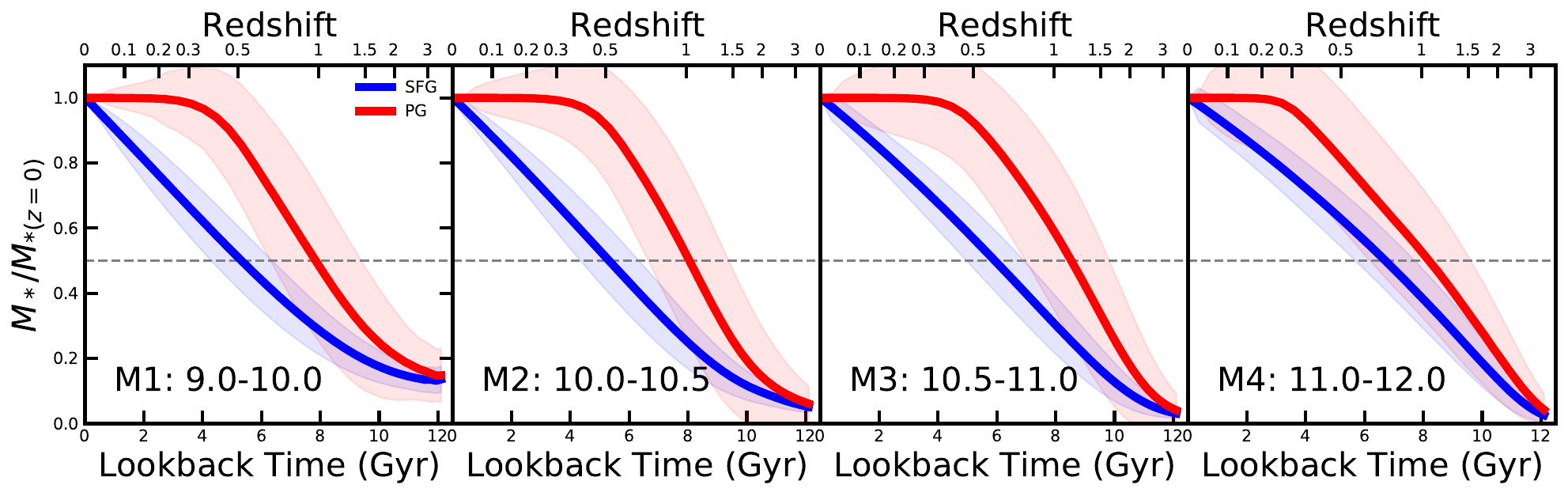}
		\caption{Stellar mass growth history of star-forming central galaxies (SFGs, blue) and passive central galaxies (PGs, red) in L-GALAXIES, normalized to the stellar mass at $z=0$. Each panel corresponds to a different range of present-day stellar mass: $\log(M_*/\mathrm{M_\odot})=9.0-10.0$ (M1), $10.0-10.5$ (M2), $10.5-11.0$ (M3), and $11.0-12.0$ (M4), respectively. The curves depict the median values of each stellar mass bin, with shaded regions representing $1\sigma$ error on the median value. Dashed lines mark half of the present-day stellar mass. \label{MassGrowth_LG_rel}}
	\end{figure*}
	
	\begin{figure*}[htb]
		\centering
		\includegraphics[width=0.45\textwidth]{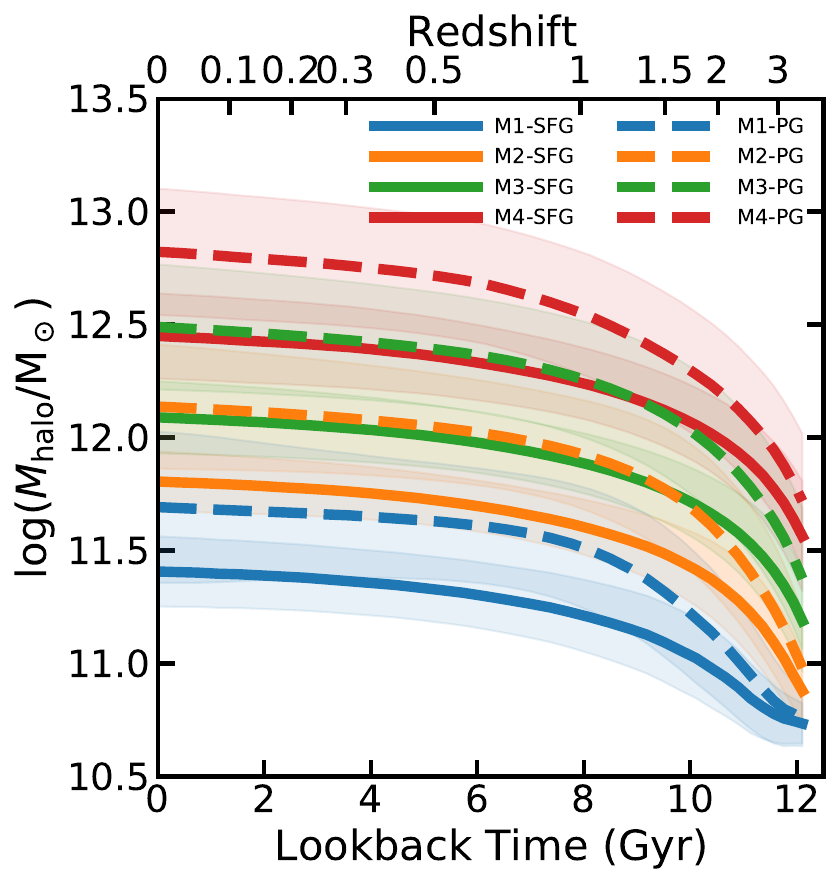}
		\includegraphics[width=0.47\textwidth]{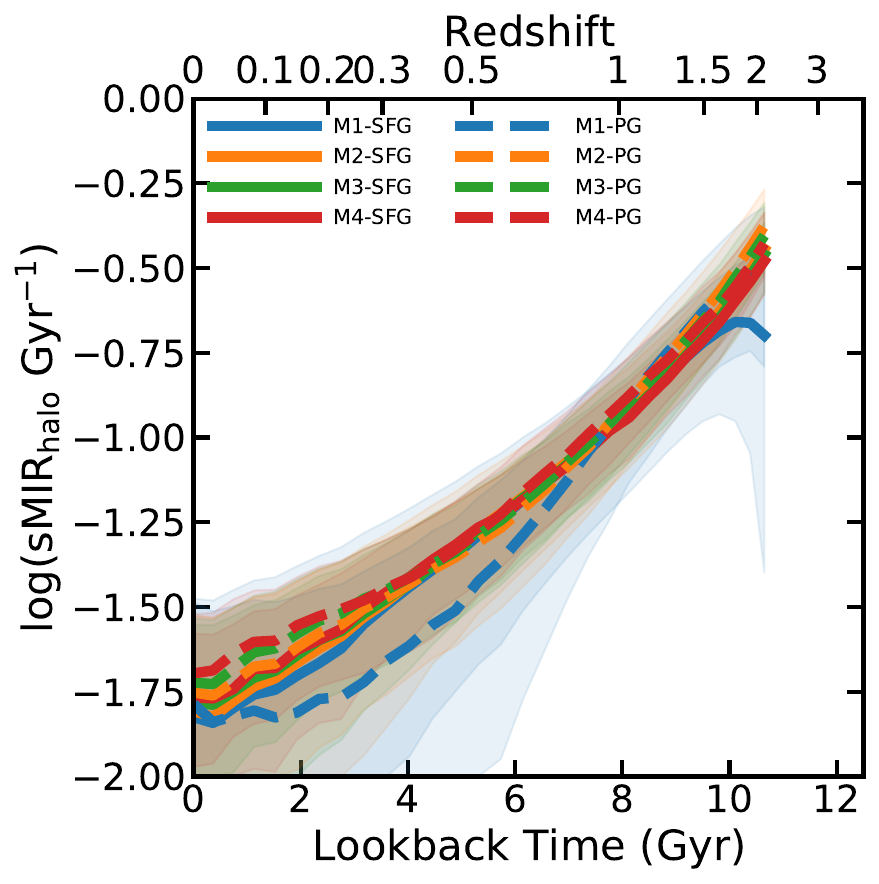}
		\caption{Halo mass assembly (left panel) and specific mass increase rate of halo ($\mathrm{sMIR_{halo}}$, right panel) as a function of the lookback time of central galaxies in L-GALAXIES. The colors represent galaxies in different stellar mass ranges at $z=0$: $\log(M_*/\mathrm{M_\odot})= 9.0-10.0$ (M1: blue), $10.0-10.5$ (M2: orange), $10.5-11.0$ (M3: green), and $11.0-12.0$ (M4: red), respectively. The solid and dashed lines show the median values of star-forming central galaxies (SFGs) and passive central galaxies (PGs), respectively. The shaded regions represent the $1\sigma$ error on the median value (left panel) or the 40th to 60th percentile variation (right panel). \label{HaloAssembly_LG_abs}}
	\end{figure*}
	
	We employ the semi-analytical model of L-GALAXIES to explore the physical origins of the observed distinct history of stellar mass assembly for SFGs and PGs. The stellar mass is defined in L-GALAXIES as the total mass of stars in the disk and the bulge. We modify the stellar mass by adding a constant factor to match the IMF used in Pipe3D catalog to ensure the consistency. To avoid systematic errors that arise from using different calibrations of the stellar mass and SFR, we redefine the SFMS in L-GALAXIES in the log(stellar mass) - log(SFR) plane:

	\begin{equation}
		\log(\mathrm{SFR}/ \mathrm{M_\odot yr^{-1}})=0.9\log(M_*/\mathrm{M_\odot})-8.8,
	\end{equation}
	\begin{equation}
	\Delta\mathrm{MS}=\log(\mathrm{SFR})-\log(\mathrm{SFR_{SFMS}}),
	\end{equation} 
	where galaxies with $\Delta\mathrm{MS}>-0.5$ are classified as SFGs, whereas those with $\Delta\mathrm{MS}<-1.5$ are classified as PGs. We have also tested alternative definition of the main sequence using a curved SFMS to define SFGs and PGs, and find small changes to the results presented in this paper.
	
	To compare with the observations, we also divide the galaxies into SFGs and PGs in four present-day stellar mass bins. Figure \ref{MassGrowth_LG_rel} illustrates the normalized history of stellar mass growth for SFGs and PGs in L-GALAXIES in each bin. Overall, The results are in qualitatively good agreement with the observed stellar mass growth history in Figure \ref{MassGrowth_MaNGA_rel}: the PGs grow faster than SFGs at high redshift and become quenched at the lookback time $\sim 2-4$ Gyr.
	
	Given the well consistent history of stellar mass growth with observations, we then utilize L-GALAXIES to explore the assembly history of dark matter halo, which is strongly connected with the growth of baryon \citep[e.g.][]{Wechsler2018} but very hard to directly observe with current tools. The host halo mass is defined as the virial mass (as defined by m\_crit200) of the FOF group the galaxy resides in. We first locate the corresponding dark matter halo for each central galaxy, and then trace the main branch of the halo merger tree. Figure \ref{HaloAssembly_LG_abs} illustrates the halo mass assembly (left panel) and specific mass increase rate of halo ($\mathrm{sMIR_{halo}}$, right panel) as a function of the lookback time of central galaxies in L-GALAXIES, categorized into various present-day stellar mass bins, denoted by different colors ($\log(M_*/\mathrm{M_\odot})= 9.0-10.0$ (M1: blue), $10.0-10.5$ (M2: orange), $10.5-11.0$ (M3: green), and $11.0-12.0$ (M4: red)). The $\mathrm{sMIR_{halo}}$ is calculated by taking the difference in halo mass between two consecutive snapshots, dividing it by the time interval, and normalizing it by the halo mass at each specific epoch.

	In the left panel of Figure \ref{HaloAssembly_LG_abs}, at a given present-day stellar mass and a given epoch, we find that PGs tend to reside in more massive halos than SFGs, which is consistent with the observation results from weak lensing \citep[e.g.][]{Mandelbaum2006, Luo2018, Bilicki2021}. This trend can be explained by the simple scenario proposed in \citet{Peng2012} and later illustrated in Figure 2 of \citet{Man2019}. For the star-forming galaxy, both the central galaxy and its halo exhibit continuous mass growth over time. While for the present-day quenched central galaxy, its star formation quenched at high redshift (due to whatever reason). In this case, its stellar mass remains about constant unless further mergers occur, while its halo continues to grow through merging with smaller halos, regardless of the star formation status of the central galaxy. This scenario produces an observed trend that at a given stellar mass, PGs reside in more massive halos than SFGs. In the right panel of Figure \ref{HaloAssembly_LG_abs}, on average, we find the specific mass increase rate of halo ($\mathrm{sMIR_{halo}}$) exhibits a weak dependence on halo mass. This weak dependence is qualitatively consistent with the results derived from cosmological hydrodynamic simulations. For instance, \cite{Faucher-Giguere2011} demonstrated a power-law relation of $\mathrm{sMIR_{halo}}$ $\sim$ $M_\mathrm{halo}^{0.06}$ with respect to halo mass. As also discussed in \citet{Dekel2013}, the weak dependence of $\mathrm{sMIR_{halo}}$ on halo mass reflects the logarithmic slope of the fluctuation power spectrum. Consequently, there are no significant differences in $\mathrm{sMIR_{halo}}$ with time among different present-day stellar mass and star formation statuses. Therefore, the apparent divergence in halo mass growth for different stellar mass for SFGs and PGs is due to (1) different initial halo mass at high redshift (for different present-day stellar mass); (2) quenching, (at a given present-day stellar mass) which decouples the co-evolution of stellar mass and dark matter mass, as explained above.

	It is worth noting that the curves of halo assembly history are the averaged results that are smoothed over numerous galaxies in each bin. Statistically, the halo mass for central galaxies tends to increase across the main branch of the merger trees; whereas the sub-halos for satellites are subject to processes that lead to mass loss, such as tidal stripping or dynamical friction during the process of accretion \citep{vandenBosch2018}. Therefore, the halo mass of individual halo may either increase due to merger events or decrease due to stripping, and its assembly history could be highly fluctuant and not necessarily monotonically increasing over cosmic time (i.e. present-day halo mass does not necessarily equate to the maximum halo mass of a particular halo).

	Here we emphasize again that our focus is solely upon examining relative trends and scaling relations in semi-analytic model and observation. Even minor modifications to methods of measurement, such as different sets of parameters in stellar population models, could lead to significant systematic error in the final results. Moreover, it is widely recognized that L-GALAXIES simulations have limitations in reproducing the observed metallicity evolution, in particular at high redshift, as pointed out by previous studies \citep[e.g.][]{Yates2012}. For instance, it fails to exhibit clear metallicity evolution in neutral gas due to the over-retention of newly synthesized metals in the interstellar medium during the early stages \citep{Yates2021}.
	
	\subsection{Halos assembled earlier or later}
	\begin{figure*}[htb]
		\centering
		\includegraphics[width=1\textwidth]{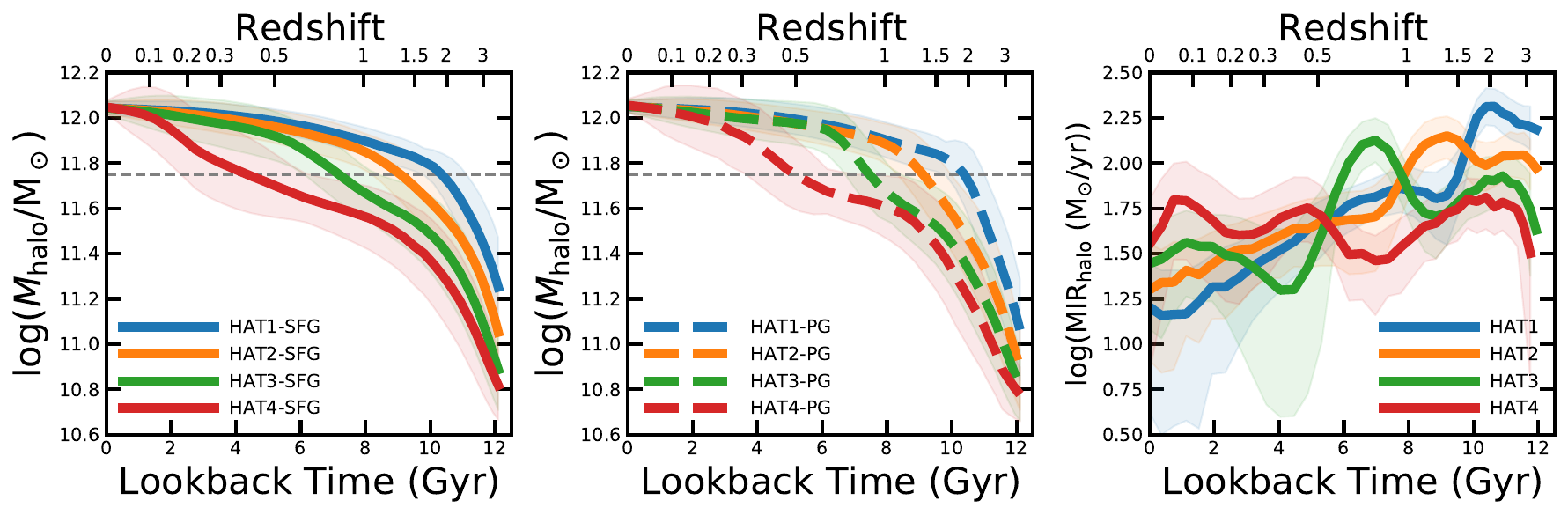}
		\caption{Left and middle panels: Halo mass assembly as a function of the lookback time of the four subsamples at a given halo mass ($M_{\mathrm{halo}}=10^{12.0}-10^{12.1}\mathrm{M_\odot}$) at $z=0$, binned by halo assembly time ($t_{\mathrm{h,50}}$, defined as the lookback time of a halo first assembled half of the present-day halo mass) for star-forming central galaxies (left panel) and passive central galaxies (middle panel) in L-GALAXIES. These halos are divided into four subsamples according to their $t_{\mathrm{h,50}}$ value, marked by blue (HAT1, $t_{\mathrm{h,50}}=10-12$ Gyr), orange (HAT2, $t_{\mathrm{h,50}}=8-10$ Gyr), green (HAT3, $t_{\mathrm{h,50}}=6-8$ Gyr), and red (HAT4, $t_{\mathrm{h,50}}<6$ Gyr) curves. The shaded regions represent the $1\sigma$ error on the median value and the gray dashed horizon line represents the average value of half present-day halo mass. Right panel: Mass increase rate of halo ($\mathrm{MIR_{halo}}$) as a function of the lookback time for the same four subsamples of different $t_{\mathrm{h,50}}$ in L-GALAXIES, with the shaded regions covering 40th to 60th percentiles variation. \label{Halo12_Th50_HaloAssembly}}
	\end{figure*}
	
	\begin{figure}[htb]
		\centering
		\includegraphics[width=0.45\textwidth]{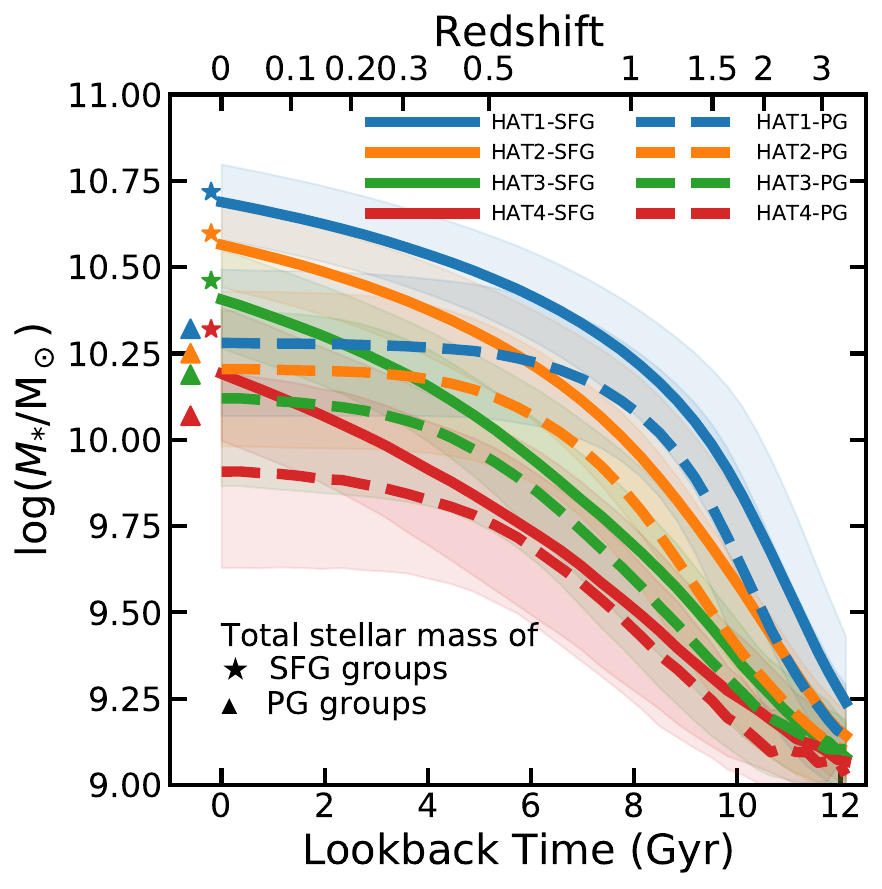}
		\caption{Stellar mass growth history of the eight subsamples of central galaxies residing in halos of different $t_{\mathrm{h,50}}$ at a given halo mass ($M_{\mathrm{halo}}=10^{12.0}-10^{12.1}\mathrm{M_\odot}$) at $z=0$ in L-GALAXIES. Each color represents a different $t_{\mathrm{h,50}}$ bin, consistent with those in Figure \ref{Halo12_Th50_HaloAssembly}. The solid and dashed lines denote the median values of star-forming central galaxies (SFGs) and passive central galaxies (PGs), respectively, while the shaded regions indicate the $1\sigma$ error on the median value. The ``stars" and ``triangles" show the median values of the present-day total stellar mass for groups with a star-forming or passive central galaxy, respectively, and are randomly offset to prevent overlap with other curves. \label{Halo12_Th50_MassGrowth}}
	\end{figure}

	We further explore in semi-analytic model to gain a better understanding of how the galaxy growth is connected to the halo assembly history. Theoretical studies suggest that in addition to halo mass, secondary properties such as halo assembly time can have a significant impact on galaxy assembly \citep[e.g.][]{Wechsler2018, Behroozi2019}. However, since halo assembly can hardly be directly constrained with current observations, we first investigate this effect using semi-analytic model. Specifically, we select galaxies of halo mass in the range ($10^{12.0}-10^{12.1}\mathrm{M_\odot}$) at $z=0$ in L-GALAXIES, which is around the typical halo mass of Milky Way-like galaxies. At this given present-day halo mass, we divide these galaxies into four subsamples based on the value of Halo Assembly Time $t_{\mathrm{h,50}}$, defined as the lookback time of a halo first assembled half of the present-day halo mass (HAT1: $10-12$ Gyr; HAT2: $8-10$ Gyr; HAT3: $6-8$ Gyr; HAT4: $<6$ Gyr). The values of $t_{\mathrm{h,50}}$ in L-GALAXIES are calculated by linear interpolation between two lookback times that include the time at which 50\% of the present-day halo mass was assembled. Although L-GALAXIES only provides halo mass at a limited number of snapshots, we are able to derive the values of $t_{\mathrm{h,50}}$ through this interpolation procedure.
	
	The left and middle panels of Figure \ref{Halo12_Th50_HaloAssembly} show the halo assembly history of the SFG hosts and PG hosts in four subsamples. For the halos that assembles earlier (HAT1), their halo mass rapidly reaches $t_{\mathrm{h,50}}$ at $z\sim 2$ and then evolves slowly. Conversely, for the halos that assembles later (HAT4), their halo mass experiences a long fast-growth phase until $z\sim0.3$. The right panel of Figure \ref{Halo12_Th50_HaloAssembly} shows the mass increase rate of halo of the four subsamples. Generally, the mass increase rate of halo decreases with cosmic time for each subsample, albeit with episodes of rapid increase showing as significant bumps on the overall decreasing curves. It is worth noting that the unsmooth curves and bumps are caused by the binning criteria based on the absolute $t_{\mathrm{h,50}}$ values, as the distribution of $t_{\mathrm{h,50}}$ is not uniform across lookback time.
	
    We examine the galaxy growth history of present-day SFGs and PGs residing in halos of different assembly times at a given present-day halo mass, shown in Figure \ref{Halo12_Th50_MassGrowth}. We also mark the median values of the present-day total stellar mass of groups with a star-forming/passive central galaxy by ``stars''/``triangles''. This figure reveals five interesting results: (1) Different halo assembly history produces very different final stellar mass of the central galaxies within, and the difference in the present-day stellar mass can be larger than $\sim 0.8$ dex when all centrals are considered (between the blue solid line and the red dashed line), or $\sim 0.5$ dex (between the blue solid line and the red solid line) for SFGs and $\sim 0.4$ dex (between the blue dashed line and the red dashed line) for PGs. This suggests halo assembly time has an important impact on galaxy growth, for both SFGs and PGs; (2) At the same present-day halo mass, halo with a higher $t_{\mathrm{h,50}}$ value (i.e. assembled earlier) hosts a more massive central galaxy than halo with a lower $t_{\mathrm{h,50}}$ value (i.e. assembled later), for both SFGs and PGs. This is due to the on average higher molecular gas fraction and higher star formation efficiency at earlier epoch \citep{Tacconi2018, Tacconi2020}. The central galaxy in the earlier assembled halo hence catches the golden times for rapid star formation in the early universe; (3) PGs feature two-phase stellar mass growth history in which they actively form stars at early times, and then cease growth after quenching and show plateaued assembly curves. At a given present-day halo mass, on average, the galaxies residing in early-assembling halos quench earlier than those in halos that assemble later, indicated by the turning point (i.e. the ``knee'') of each curve. This is consistent with the results shown in Figure \ref{Halo12_Th50_HaloAssembly}, where galaxies residing in early-assembling halos reach higher halo mass than those in late-assembling halos at a given epoch; (4) At a given present-day halo mass and a given assembly time, SFGs have a higher present-day stellar mass than PGs. This may be due to the decoupling of halo assembly and galaxy growth after the quenching of star formation which ceases stellar mass increase but cannot stop halos from further accretion; (5) On average, the stellar mass difference between centrals and the groups (i.e. the total stellar mass of the satellites), is smaller in halos that assembled earlier. This may be explained by that central galaxies in early-assembling halos have more time to gather as many baryons in the group as possible through mergers and interactions, which results in lower luminosity/mass gaps. We plan to derive halo assembly time in observation in our future work to test these predictions. These trends also hold when we study halos of lower ($10^{11.5}-10^{11.6}\mathrm{M_\odot}$) and higher ($10^{12.5}-10^{12.6}\mathrm{M_\odot}$) mass, respectively.
	
\begin{figure}[htb]
		\centering
		\includegraphics[width=0.45\textwidth]{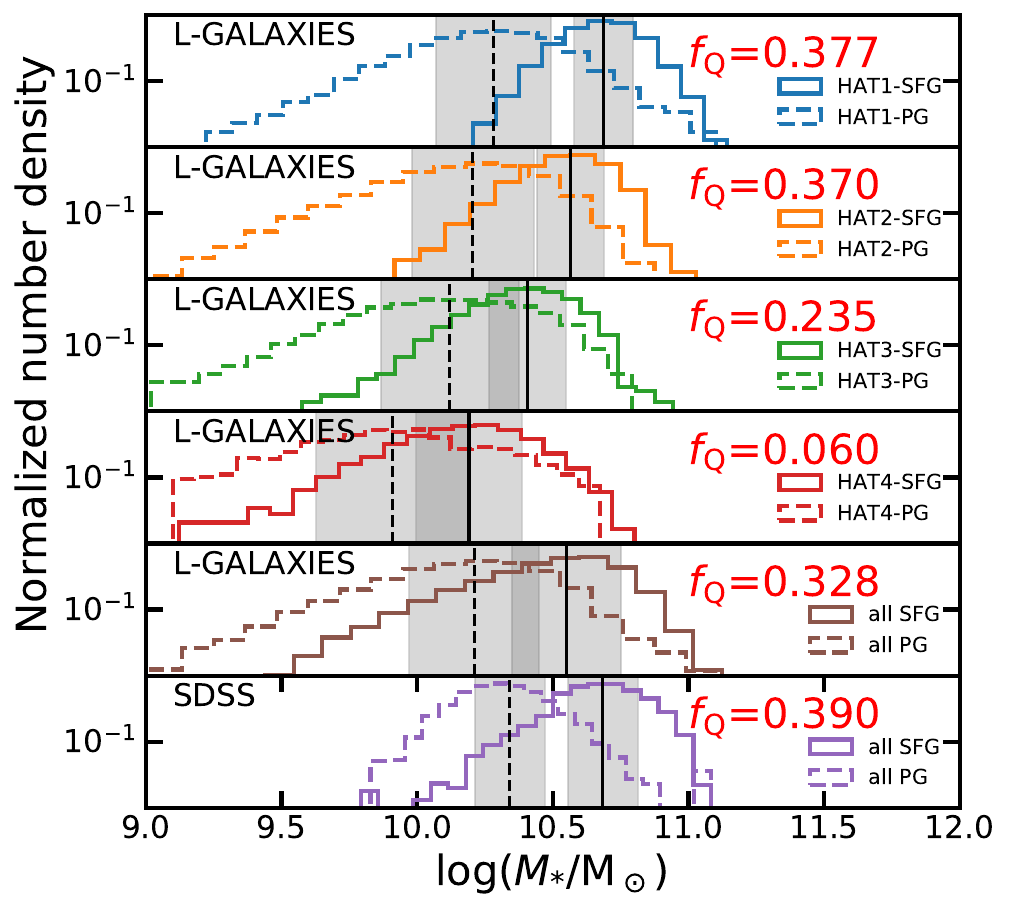}
		\caption{Top four panels: Stellar mass distribution of star-forming central galaxies (SFGs) and passive central galaxies (PGs) residing in halos of different $t_{\mathrm{h,50}}$ at a given halo mass ($M_{\mathrm{halo}}=10^{12.0}-10^{12.1}\mathrm{M_\odot}$) at $z=0$ in L-GALAXIES. These colors indicate different $t_{\mathrm{h,50}}$ bins, consistent with Figure \ref{Halo12_Th50_HaloAssembly}. The fifth panel: Stellar mass distribution of all SFGs and PGs residing in $M_{\mathrm{halo}}=10^{12.0}-10^{12.1}\mathrm{M_\odot}$ halos. Each panel presents two histograms, one for SFGs and one for PGs, plotted by solid and dashed step lines, respectively. The vertical black solid/dashed lines represent the median stellar mass of SFGs/PGs at $z=0$, and the shaded regions indicate $1\sigma$ error on the median value. The quenched fraction for each panel is given in the top-right corner, calculated with a stellar mass cut of $M_{*} \ge 10^{10.0} \mathrm{M_\odot}$. Bottom panel: The same as other panels but for central galaxies ($z=0.02-0.10$) residing in $M_{\mathrm{halo}}=10^{11.9}-10^{12.1}\mathrm{M_\odot}$ halos in SDSS DR7 with $V_{\mathrm{max}}$ weighted. The halo mass of central galaxies in SDSS DR7 is derived from Zhao et al. (to be submitted).
		\label{Halo12_SFGPGdistribution_LGMANGA}}
	\end{figure}

	As a result, the present-day stellar mass distributions of central galaxies are significantly different for halos that assembled at different time, as shown in Figure \ref{Halo12_SFGPGdistribution_LGMANGA}. The top four panels reveal a clear trend in L-GALAXIES, that galaxies are on average less massive when the halos are assembled at a later time. Furthermore, at a given halo assembly time, the median stellar mass of SFGs is higher than that of PGs, consistent with the results in Figure \ref{Halo12_Th50_MassGrowth}. This trend can be explained that at a given halo mass, the total amount of baryon is fixed, and present-day SFGs are more efficient at converting baryons into stellar mass than PGs. Consequently, SFGs exhibit a higher peak in the present-day stellar masses distribution on average. This result can be understood by the distinct scaling relations for SFGs and PGs in the $\log M_{*}$-$\log M_{\mathrm{h}}$ plane \citep[e.g.][]{Man2019}, and is also consistent with the results from independent measurements by weak lensing \citep[e.g.][]{Mandelbaum2006,Luo2018,Bilicki2021}. Notably, the difference in stellar mass distribution between SFGs and PGs reduces from early to late halo assembly time, as we move from top to bottom panels (of the top four panels), and the quenched fraction of central galaxies (defined as the number ratio of quenched centrals to all centrals with a stellar mass cut of $M_{*} \ge 10^{10.0} \mathrm{M_\odot}$ and quenched galaxies are those with SFR one dex lower than galaxies on the SFMS) also decreases. In fifth panel of Figure \ref{Halo12_SFGPGdistribution_LGMANGA}, we plot the present-day stellar mass distribution of all galaxies in this halo mass range in L-GALAXIES. 
	
\begin{figure*}[htb]
		\centering
		\includegraphics[width=0.45\textwidth]{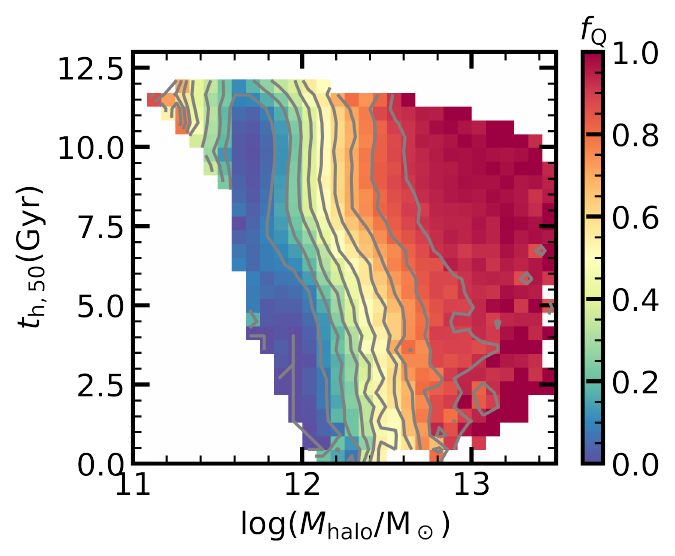}
		\includegraphics[width=0.47\textwidth]{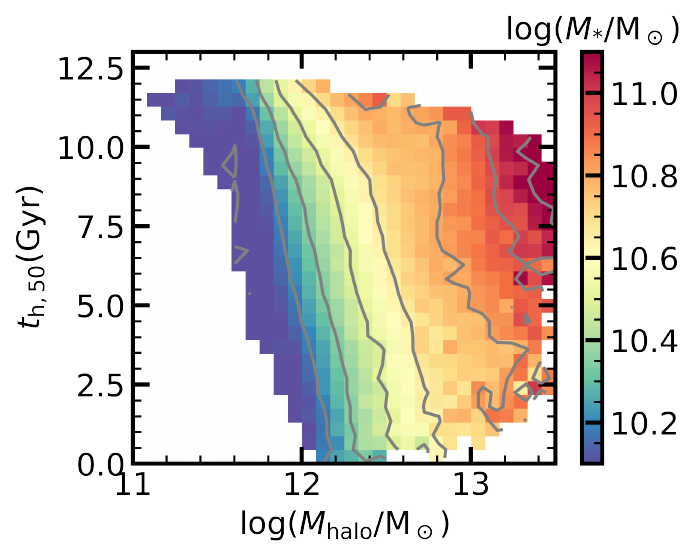}
		\caption{Quenched fraction (left panel) and averaged stellar mass (right panel) of central galaxies as a function of halo mass and halo assembly time $t_{\mathrm{h,50}}$ in L-GALAXIES, with a stellar mass cut of $M_{*} \ge 10^{10.0} \mathrm{M_\odot}$. The red/blue color codings indicate high/low value of quenched fraction (left panel) and averaged stellar mass (right panel). \label{Fq_LG}}
	\end{figure*}

	The left panel of Figure \ref{Fq_LG} presents the quenched fraction of central galaxies as a function of halo mass and halo assembly time $t_{\mathrm{h,50}}$ in L-GALAXIES, with a stellar mass cut of $M_{*} \ge 10^{10.0} \mathrm{M_\odot}$. For comparison, we also show the quenched fraction of central galaxies with the stellar mass limits of $M_{*} \ge 10^{9.0} \mathrm{M_\odot}$ and $M_{*} \ge 10^{9.5} \mathrm{M_\odot}$ in Appendix \ref{app}. The sharp gradient of quenched fraction with respect to halo mass suggests that the halo potential plays a dominant role in star formation quenching of central galaxies, while the still significant gradient along the axis of halo assembly time indicates the non-negligible influence from the assembly time of halos (particularly for central galaxies residing in $10^{11.8}-10^{12.8}\mathrm{M_\odot}$ halos). At the same present-day halo mass, the halos assembled earlier have a higher averaged quenched fraction of the centrals than those assembled later. This can be partly explained by the correlation, shown in the right panel of Figure \ref{Fq_LG}, that halos assembled earlier host more massive centrals thus with a higher quenched fraction, in particular around the ``golden halo mass'' around $10^{12.0}\mathrm{M_\odot}$ as proposed in \citet{Dekel2019}.
	
	In observation, at a given present-day halo mass, the $V_{\mathrm{max}}$-weighted stellar mass distribution difference between SFGs and PGs is also remarkable and qualitatively consistent with the peaks in L-GALAXIES, shown in the bottom panel of Figure \ref{Halo12_SFGPGdistribution_LGMANGA} for similar halo mass range ($10^{11.9}-10^{12.1}\mathrm{M_\odot}$) at $z=0.02-0.10$ in SDSS DR7 sample. The halo mass is from the catalog of Zhao et al. (to be submitted) which derives halo mass from a series of galaxy observables based on the models established in L-GALAXIES using machine learning techniques. The derived halo mass function of Zhao et al. (to be submitted) nicely recovers the theoretical halo mass function and their derived stellar-to-halo mass relation, separately for SFGs and PGs, is consistent with independent measurements by weak gravitational lensing \citep[e.g.][]{Mandelbaum2006, Luo2018, Bilicki2021}. The $V_{\mathrm{max}}$ values are derived from k-correction program v4\_1\_4 \citep{Blanton2007}, to account for any residual volume incompleteness within this redshift bin. The absence of galaxies in the low-mass end is due to incompleteness in observations, thus we calculate the quenched fraction with a stellar mass cut of $M_{*} \ge 10^{10.0} \mathrm{M_\odot}$. In contrast to L-GALAXIES, we do not divide these galaxies by different $t_{\mathrm{h,50}}$ values, but we plan to explore the effect of different halo assembly history in observation in future work.
	
	Moreover, at a given halo mass and a given halo assembly time, there exists both SFGs and PGs with significant variations in present-day stellar mass. However, it should be noted that the halo assembly time merely acts as a proxy and cannot fully reflect the complete halo assembly history. Additionally, these two halo properties alone cannot entirely determine the present-day status of a central galaxy. Baryon processes including feedback, also play a crucial role. It is worth noting that our results are based on the treatment of astrophysical processes in L-GALAXIES (see Section \ref{LG}). In fact, the specific quenching mechanisms in L-GALAXIES are complex, involving both dark matter halo and AGN, which is beyond the scope of this work. Moreover, different semi-analytic models and simulations employ different recipes, particularly regarding feedback mechanisms. One may suspect that the results presented in this work may be dependent on the choice of the semi-analytic model. L-GALAXIES has been calibrated and tuned to have well-reproduced stellar mass function and quenched fraction across various redshifts in observations. This puts strong constraints on the stellar mass growth history and quenching history. Hence we would expect that alternative semi-analytic models or simulations capable of reproducing the observed stellar mass function and quenched fraction would yield similar stellar mass growth history and quenching history, even if they employ different quenching mechanisms. It is worth emphasizing that the scientific goal of this paper is to highlight the strong impact of halo mass and its assembly history on galaxy properties, not to explore the detailed star formation and quenching physics (which is the goal in our following work). Hence the logical choice is to use a semi-analytic model/simulation that can best match the observed stellar mass function and quenched fraction (even the physics inside the semi-analytic model/simulation may not be correct). In the future, we will compare these trends across different semi-analytic models and simulations, and investigate the relationship between the physical recipes and the present-day star formation status.
	
	\section{Discussion}\label{discussion}
	
	\subsection{Halo assembly and galaxy growth}
	In this work, we investigate the connections between stellar mass growth history, quenching history and halo assembly history for central galaxies in both observation and semi-analytic model. Our analysis reveals that more massive galaxies and those residing in more massive halos tend to assemble their stellar mass earlier, possess a larger stellar mass, and quench earlier than their less massive counterparts inhabiting less massive halos. These trends have been previously found in various works \citep[e.g.][]{Cowie1996, Conroy2009, Fontanot2009, Dave2016, Chaves-Montero2020} and confirmed through multiwavelength observations of galaxy populations at different redshifts \citep[e.g.][]{Pozzetti2010, Leitner2012, Bauer2013, Tomczak2016}.
	
	Our results provide compelling evidence for the intricate link between galaxy growth and halo assembly history. Specifically, at a given present-day halo mass, we find that central galaxies residing in early-assembling halos generally grow earlier, quench earlier, and attain a larger present-day stellar mass than those in late-assembling halos. This trend is consistent with numerous previous works \citep[e.g.][]{Ibarra-Medel2016, Martin-Navarro2022}. Such link has also been studied by other theoretical works, e.g. using EAGLE hydrodynamic simulation \citep{Matthee2017} and semi-analytic models \citep{Croton2007, Tojeiro2017}. They showed that early-forming halos typically accrete a higher fraction of their baryons (i.e. higher $M_*/M_{\mathrm{halo}}$) at early times and this leads to the formation of more massive galaxies than late-forming halos of the same present-day halo mass. However, some works predicted that the baryon conversion efficiency reaches its maximum at late times when late-forming halos are accreting \citep[e.g.][]{Behroozi2013, Moster2013}. This implies that late-forming halos should have a higher stellar mass than early-forming halos, which contradicts the previous results. Therefore, further analysis is required from both the theoretical and observational perspectives to reconcile this situation \citep{Wechsler2018}. For a galaxy with $M_{\mathrm{halo}} \sim 10^{12} \mathrm{M_\odot}$, our analysis reveals that central galaxies residing in early-assembling halos tend to quench earlier, resulting in a larger quenched fraction. This trend implies that the rapid or early accretion of halos may drive the quenching of central galaxies. However, \citet{Tinker2018} argued that the quenching processes have limited correction with halo formation history.
	
	We also propose that halo assembly time ($t_{\mathrm{h,50}}$) could serve as a secondary parameter of halo assembly bias. This bias has been extensively studied in the literature, with many halo properties serving as proxies, such as concentration, spin, etc. \citep{Wechsler2001, Gao2005, Wechsler2006}. \citet{Montero-Dorta2021} suggested using the slope ($\beta$) of the specific mass accretion history of the hosting halos (sMAH) as an indicator of halo assembly bias. They found in TNG300 that halos with increasingly steeper $\beta$ grew more massive earlier, with their hosting galaxies exhibiting larger stellar-to-halo mass ratios, faster gas depletion, earlier quenching, and peak star formation history at higher redshifts. In our future work, we will also explore how to measure halo assembly history in observation.
	
	\subsection{Caveats of fossil record method}
	
	In this work, we conduct a comparison between observational data in SDSS-MaNGA and data in semi-analytic model L-GALAXIES, and find some differences in the absolute values. From the standpoint of observation, the primary source of uncertainty lies in the fossil record method employed by Pipe3D. While it can accurately reproduce the details of the observed spectra, it is prone to significant uncertainties \citep[e.g.][]{CidFernandes2014, Sanchez2016, Sanchez2020}. In contrast to assuming a parameterized star formation history, this algorithm uses combinations of stellar ages and metallicity from different lookback times for each galaxy to fit the luminosity evolution obtained from spatially resolved IFS observations. As a result, there may be inconsistencies in the stellar mass growth history and chemical enrichment history of each galaxy. Furthermore, the results generated by Pipe3D are highly dependent on the selected templates of SSPs or the stellar libraries utilized \citep[e.g.][]{GonzalezDelgado2014}. The uncertainties associated with the results also encompass factors such as galaxy inclination, spatial distributions, dithering and spatial binning procedures, as well as the signal-to-noise ratio (SNR) of the spectra \citep[for detailed discussion, see e.g.][]{Sanchez2016, Ibarra-Medel2016, Ibarra-Medel2019}.
	
	The limitations discussed above can significantly impact the stellar mass growth history. For instance, the shape of the reconstructed SFH can become markedly smoothed (or flattened) at old ages due to the coarse SSP age binning employed, which is particularly inadequate for ages exceeding $\sim$ 8 Gyr. The process of deducing stellar metallicity, specifically the absolute value, is highly dependent on the data utilized, calibrated, and the methodology employed to examine the stellar populations \citep{Sanchez2020}. The inferred stellar metallicity enrichment history may be biased, particularly at higher redshifts. For example, as shown in Figure \ref{Zenrich_MaNGA_abs_agebin}, the stellar metallicity of passive galaxies already surpasses that of star-forming galaxies at the early epoch ($z\sim2$), and this can be partly explained that the features of star formation from more than 10 Gyr ago are challenging to be observed and inferred.
	
	\section{Conclusions}\label{conclusion}
	
	In this work, we first investigate the stellar mass growth history and chemical enrichment history of central galaxies in MaNGA survey. Combining these observations and results from L-GALAXIES semi-analytic model, we find that the general trend of observational derived stellar mass growth history is in qualitatively good agreement with that in semi-analytic model, for both star-forming and passive central galaxy populations, and that both halo mass and halo assembly history have significant impacts on the stellar mass growth history and star formation quenching history of the central galaxies. Our main results are summarized as follows:
	
	(1) At a given epoch, the derived stellar metallicity of MaNGA passive central galaxies is always higher than that of the star-forming ones, and this stellar metallicity difference is larger for low-mass galaxies. While at a given stellar mass, the stellar metallicity difference becomes increasingly larger at an earlier epoch (Figure \ref{Zenrich_MaNGA_abs_agebin} and Figure \ref{MZRdifference_MaNGA}). \citet{Peng2015} showed that during strangulation the amount of stellar metallicity enhancement is determined by the gas fraction of the galaxies (and also the strangulation timescale). The observed gas fraction becomes larger at lower stellar masses (at a given epoch) and at an earlier epoch (at a given stellar mass), therefore these stellar metallicity difference trends shown in Figure \ref{Zenrich_MaNGA_abs_agebin} are consistent with the strangulation scenario. This implies that strangulation is the primary mechanism for shutting down star formation in central galaxies not only in the local universe as found in \citet{Peng2015}, but also very likely for central galaxies at higher redshifts. As shown in Figure \ref{MZRdifference_MaNGA}, at high-mass end, compare to $z\sim0$, the stellar metallicity difference between passive central galaxies and star-forming central galaxies at high redshifts is not notably different while the gas fraction is significantly higher. This suggests that for the most massive galaxies at $z\sim2$, gas removal via feedback is still required in addition to strangulation due to their significantly higher gas fraction compared to their local counterpart. In addition, the physical mechanism of strangulation from $z\sim0$ to $z\sim2$ might be closely related to the weakly mass-dependent characteristic $\Sigma_{1\ \mathrm{kpc}}^{crit}$ (above which quenching is operating) for central galaxies as found in the local universe \citep{Xu2021} and in higher redshifts to $z\sim2.5$ \citep{Xu2023}.
	
	(2) At the same present-day stellar mass, the star-forming central galaxies and passive central galaxies have distinct stellar mass growth history. On average, passive central galaxies assembled half of their final stellar mass $\sim2$ Gyr earlier than star-forming central galaxies (Figure \ref{MassGrowth_MaNGA_rel}), except for the low-mass galaxies which show no significant difference, probably due to their more bursty star formation history and more frequent rejuvenation. These observational derived trends are in qualitatively good agreement with results in semi-analytic model, for both star-forming and passive galaxy populations.

	(3) To understand the observed stellar mass assembly history difference for star-forming and passive centrals, we explore in the semi-analytic model and find that at a given present-day stellar mass, the passive central galaxies reside in, on average, more massive halos with a higher halo mass increase rate across cosmic time (Figure \ref{HaloAssembly_LG_abs}). As a consequence, passive central galaxies are assembled faster and also quenched earlier than their star-forming counterparts.
	
	(4) To better understand the connections between galaxy growth history and halo assembly history, we further explore in semi-analytic model the role of halo assembly history in driving galaxy assembly. As shown in Figure \ref{Halo12_Th50_MassGrowth}, we find that:
	\begin{itemize}
	\item Different halo assembly history produces very different final stellar mass of the central galaxies within, the difference in the present-day stellar mass can be larger than $\sim 0.8$ dex when all centrals are considered, or $\sim 0.5$ dex for star-forming galaxies and $\sim 0.4$ dex for passive galaxies.
	\item At the same present-day halo mass, halo with a larger $t_{\mathrm{h,50}}$ (i.e. assembled earlier) host a more massive central galaxy than halo with a smaller $t_{\mathrm{h,50}}$ (i.e. assembled later). This is due to the on average higher molecular gas fraction and higher star formation efficiency at earlier epoch. The central galaxy in the earlier assembled halo hence catches the golden times for rapid star formation in the early universe. 
	\item For halo with the same $t_{\mathrm{h,50}}$, the star-forming central galaxy on average is more massive than the passive ones. This is simply because for halos with the same $t_{\mathrm{h,50}}$, they have similar halo accretion history and hence similar baryon content at a given epoch. The central  star-forming galaxy by definition has converted more baryons into stars through star formation than the passive one (quenched due to whatever reason).
	\end{itemize}
	
	(5) Finally, we show in Figure \ref{Halo12_SFGPGdistribution_LGMANGA} the distribution of the stellar mass of star-forming central galaxies and passive central galaxies at a given present-day halo mass of $\sim10^{12} \mathrm{M_\odot}$, in both L-GALAXIES and SDSS. At the same present-day halo mass, the halos assembled earlier have a higher averaged quenched fraction of the centrals than those assembled later. This is likely due to the fact that halos assembled earlier on average host more massive central galaxies. Figure \ref{Fq_LG} further shows in a more clear way that in L-GALAXIES, halos assembled earlier host more massive centrals with a higher averaged quenched fraction, in particular around the ``golden halo mass" around $10^{12} \mathrm{M_\odot}$ as proposed in \citet{Dekel2019}.
	
	Our result that strangulation is also the primary mechanism for star formation quenching in central galaxies at higher redshifts is based on the chemical enrichment history derived from MaNGA spectrum in the local universe, hence is not from direct measurement. The forthcoming Main MOONS GTO Extragalactic Survey on VLT \citep[MOONRISE,][]{Maiolino2020} will directly measure the stellar metallicity difference between star-forming and passive galaxies at $z>1$, and test our result.  In the meanwhile, we show that both halo mass and halo assembly history have significant impacts on the stellar mass growth history and star formation quenching history of the central galaxies. These results call attention back to the dark matter halo as an important driver, or even the dominant driver of the galaxy evolution, comparing to baryonic processes such as AGN feedback. In Zhao et al. (to be submitted), they have accurately measured the halo mass of individual SDSS groups, and the resulted stellar-to-halo mass relation (SHMR) of star-forming centrals and passive centrals now match those measured from weak lensing perfectly. In our future work, we will also measure the halo assembly history in observation. Together with the halo mass and halo assembly history for SDSS groups, it will allow us to better explore the connections between stellar mass growth history, quenching history and halo assembly history.
	
	\begin{acknowledgments}
		We thank the anonymous referee for useful comments that have improved the paper. YJP and CQL acknowledge the support by the National Science Foundation of China (NSFC) Grant No. 12125301, 12192220, 12192222, and the science research grants from the China Manned Space Project with NO. CMS-CSST-2021-A07. LCH acknowledges the support by the National Science Foundation of China (11721303, 11991052, 12011540375, 12233001), the National Key R\&D Program of China (2022YFF0503401), and the China Manned Space Project (CMS-CSST-2021-A04, CMS-CSST-2021-A06).
	\end{acknowledgments}

	
	\bibliography{halo_ref}{}    
	\bibliographystyle{aasjournal}  
	\appendix

	\section{Quenched fraction on different stellar mass cuts}\label{app}
	\renewcommand{\appendixname}{Appendix~\Alph{section}}
	\setcounter{figure}{0} 
	\renewcommand{\thefigure}{A\arabic{figure}}
	
\begin{figure}[htb]
		\centering
		\includegraphics[width=0.45\textwidth]{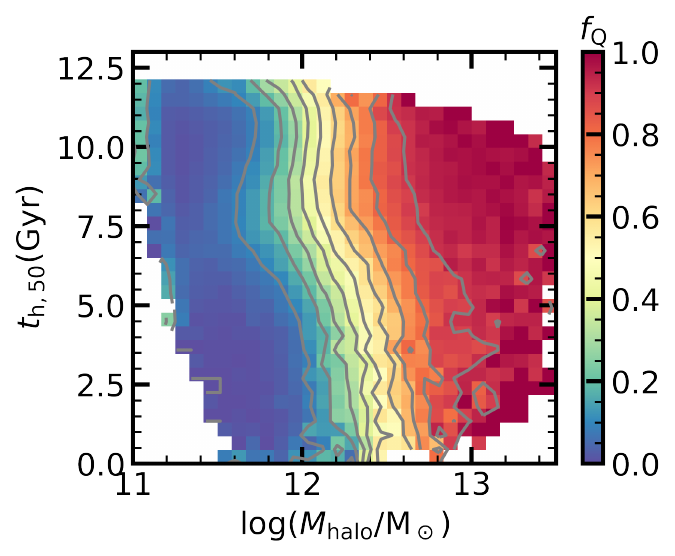}
		\includegraphics[width=0.45\textwidth]{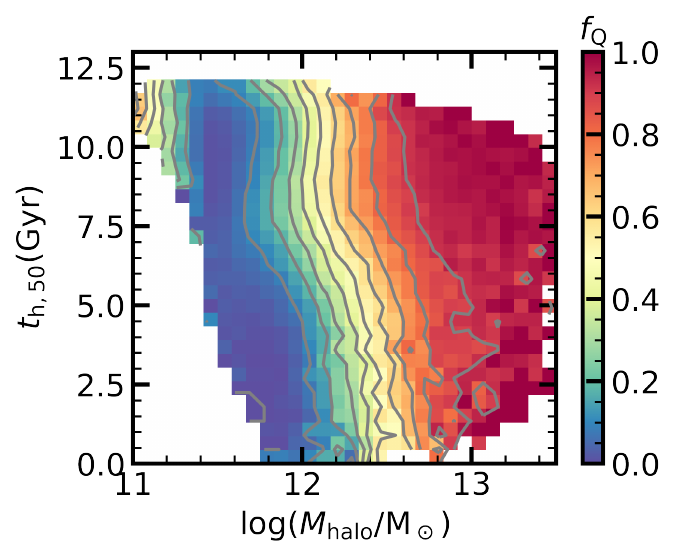}
		\caption{As for Figure \ref{Fq_LG} but with stellar mass cut of $M_{*} \ge 10^{9.0} \mathrm{M_\odot}$ (left panel) and $M_{*} \ge 10^{9.5} \mathrm{M_\odot}$ (right panel). \label{Fq_LG9095}}
	\end{figure}

	Figure \ref{Fq_LG9095} shows the quenched fraction of central galaxies as a function of halo mass and halo assembly time $t_{\mathrm{h,50}}$ in L-GALAXIES, with a stellar mass cut of $M_{*} \ge 10^{9.0} \mathrm{M_\odot}$ (left panel), and $M_{*} \ge 10^{9.5} \mathrm{M_\odot}$ (right panel). The cut at a lower stellar mass of $10^{9.0} \mathrm{M_\odot}$ extends the results to a slightly lower halo mass, and both two plots shows a similar trend to the left panel of Figure \ref{Fq_LG} at $M_{\mathrm{halo}} \ge 10^{12.0} \mathrm{M_\odot}$. This is because the stellar mass of almost all centrals residing in $M_{\mathrm{halo}} \ge 10^{12.0} \mathrm{M_\odot}$ halos is larger than $10^{10} \mathrm{M_\odot}$, hence not affected much by such a stellar mass cut.

\end{document}